# A Three-phase and Single-phase Compatible Dual-Mode EV On-Board Charger with Integrated Active Power Decoupling

Bitan Joydhar
*Department of Electrical Engineering*
*Indian Institue of Technology Kharagpur*
Kharagpur, India
bitanjoydhar25@kgpian.iitkgp.ac.in

Shimul K. Dam
*Department of Electrical Engineering*
*Indian Institue of Technology Kharagpur*
Kharagpur, India
skd@ee.iitkgp.ac.in

*Abstract*—Onboard charger (OBC) is essential part of Electric Vehicle (EV). High-performance EVs are preferring three phase charging to achieve higher power level. However, the ability to charge from a single phase supply is also required. A dualmode OBC for EVs is proposed, which is capable of operating from both three-phase supply and single-phase supply. The single phase charging comes with the requirement of bulky DC link capacitance due to double frequency current in DC link. The proposed topology eliminates this by achieving Active Power Decoupling (APD) using only one additional relay switch and a small capacitor. The proposed topology is verified under different conditions in a detailed simulation, which shows more than an order of magnitude reduction in DC link capacitance during single phase operation.

*Index Terms*—On-Board Charger, Dual-Mode OBC, Active Power Decoupling, SPWM, Open-Loop Control

## I. Introduction

The rapid adoption of electric vehicles has driven up the demand for efficient on-board chargers(OBC). This is essential because it converts AC supply to a regulated DC voltage to charge the battery. The OBCs should have high efficiency and a power factor close to unity while maintaining low size and hardware complexities.

OBCs have both two-stage and single stage power conversion method. Most of them use two-stage power conversion method [1]. In this method, the front-end part is an ACDC converter [2] and the rear-end part is an isolated DCDC converter [3]. Front-end converters act as power factor correction circuits (PFC) that regulate the DC-link voltage and ensure the grid harmonic standards compliance. The rear-end circuit is a isolated DC-DC converter such as DAB, LLC or PSFB converter [2]. The front-end converter influences the overall efficiency and power density and it is the main focus in this work. The isolated DC-DC converter part is beyond the scope of this paper.

Most of the modern OBCs are capable of using only singlephase or three-phase supplies. Single-phase OBCs are widely used in residential charging applications, but they face a double-line-frequency pulsating power that produces a 100Hz ripple on the DC link. To reduce it, Electrolytic capacitors is used, leading to decreased the power density of OBCs and short lifetime [4]. A balanced three-phase system can provide nearly constant instantaneous input power, which significantly reduces the low-frequency energy-storage requirement at the DC link. A three-phase OBC is favoured by modern highperformance EVs due to higher charging power. However, the OBC should also have single phase charging ability so that the

EV can be charged where three phase supply is not available.

Some existing OBCs have the capability of operating with both single-phase and three-phase supplies [5], but they use separate circuits for the two operating modes. This requires additional switches or auxiliary circuits to change between the two operating modes. Different modulation and control strategies are also required for the two operating modes. It is possible to use two phases of a three phase charger to charge from single phase supply. However, in single-phase operation, the same converter experiences 100-Hz pulsating power. Therefore, the DC link capacitor must be large and bulky to maintain a near-unity power factor, low THD, and stable operation in both operating modes.

Different power factor correction topologies have been proposed, such as totem-pole PFC rectifiers [1] and threephase PWM rectifiers [6]. An Active Power Decoupling (APD) method [7], [8] has also been developed to mitigate 100 Hz frequency power ripple on the DC link when operated with a single-phase supply, but it increases the hardware complexity significantly, and requires dedicated control circuits. Existing front-end converters are generally optimized for only singlephase or three-phase supply, with limited research on dualmode OBCs. Therefore, a simple and universal front-end converter is needed that can operate in both single-phase and three-phase modes without significant additional hardware complexity.

This paper proposes a dual-mode OBC that can operate with both three-phase and single-phase supplies while maintaining unity power factor and the desired DC-link voltage under the specified operating conditions. In three-phase operation, the three half-bridges function as a conventional three-phase converter, whereas in single-phase operation, two half-bridges work as a single-phase rectifier, while the third half-bridge operates as an active power decoupling circuit with additional small capacitors. Thus, using the redundant third half-bridge, APD operation is achieved during single phase charging to eliminate the need for a bulky DC link capacitor. The proposed topology allows dual supply compatibility with only an additional relay and a small capacitor and this solution can be retro-fitted to existing three phase OBCs.

## II. Proposed Topology

The proposed OBC topology is explained using its circuit diagram and equivalent circuits in the following subsections.

### A. Circuit Diagram

The circuit diagram of the proposed OBC is shown in Fig. 1. Topology is derived from both conventional three phase rectifier topology with small modification to enable Active Power Decoupling (APD) during single phase operation. The rectified dc voltage is then fed to a DC/DC converter, which charges the EV battery. This work focuses on the rectifier circuit, and the proposed modified rectifier topology can work with any of the several suitable dc-dc converter topologies.

The proposed rectifier in Fig. 1 has the same structure as a three-phase voltage source boost rectifier with a L filter at the input and a dc-link capacitor at its output. With this standard topology, an additional capacitor $C_{APD}$ is connected to one of the phase input terminals using a relay or contactor. With this simple modification, a standard three-phase rectifier can also be operated as a single-phase rectifier with APD. Moreover, this additional circuit can easily be retrofitted to an existing three-phase rectifier. The three-phase and single-phase modes of the proposed rectifier are explained in the following subsections.

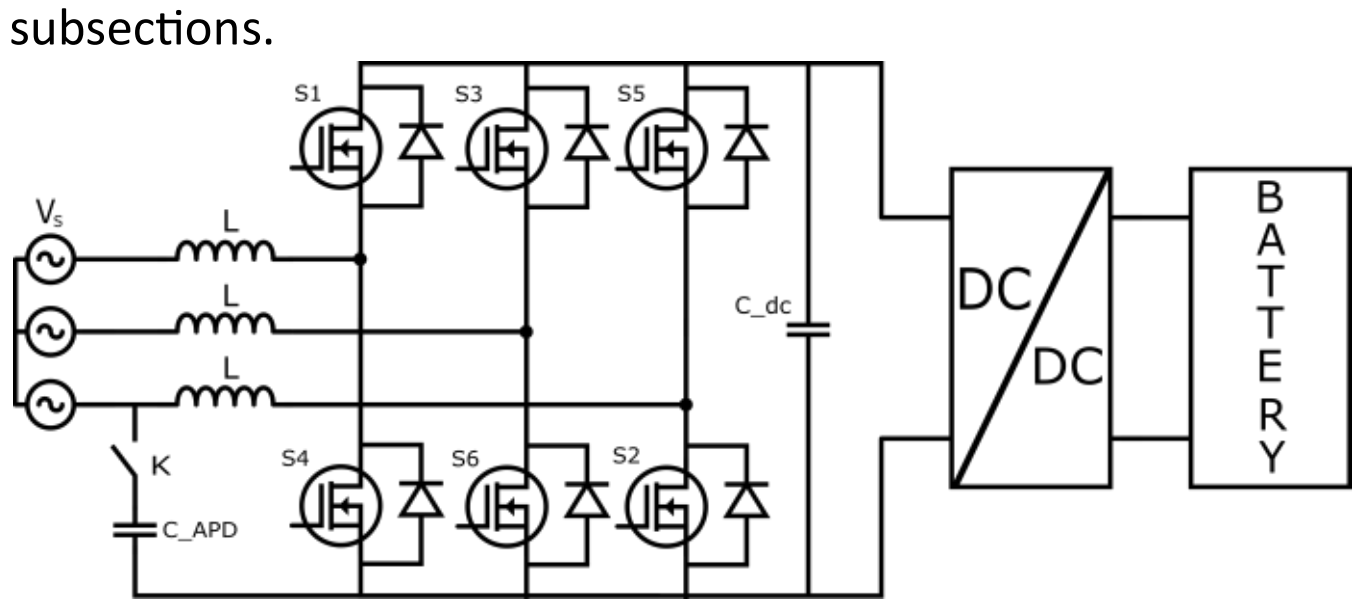

Fig. 1: Overall topology of the proposed dual-mode OBC

### B. Proposed OBC in Three-Phase Mode

In the three-phase operating mode, all three half-bridge legs of the converter are connected to a three-phase power supply. The relay switch $K$ remains open in this mode. As a result, the auxiliary APD branch is completely disconnected from the rectifier. The resulting equivalent circuit is shown in Fig. 2, which is the same as a three-phase voltage source boost rectifier. All six semiconductor switches are involved in the rectification process.

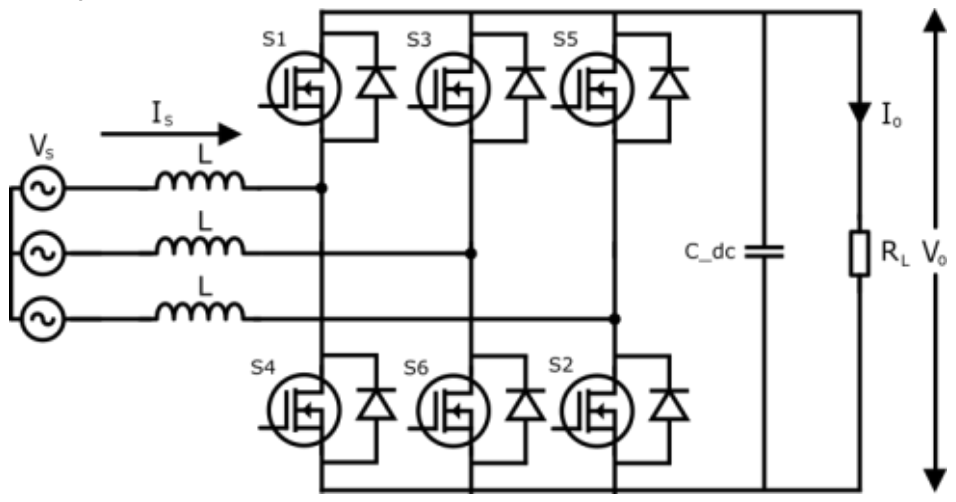

Fig. 2: Equivalent circuit of the proposed converter during three-phase operation.

### C. Proposed OBC in Single-Phase Mode

In the single-phase operating mode, two of the three converter legs are used for AC-DC conversion, forming a single-phase voltage source boost rectifier. The third converter leg, which would otherwise remain unused, is used for the power decoupling (APD) circuit. The relay switch $K$ remains closed in this mode. The third converter leg, input inductor, the relay, and the additional capacitor forms the APD circuit, as shown in Fig. 3. This circuit is used to divert the the double-line-frequency ripple energy from the DC link capacitor $C_{dc}$ to the auxiliary capacitor $C_{APD}$. Thus, the unused converter leg is utilized for active power decoupling without requiring any additional semiconductor switch.

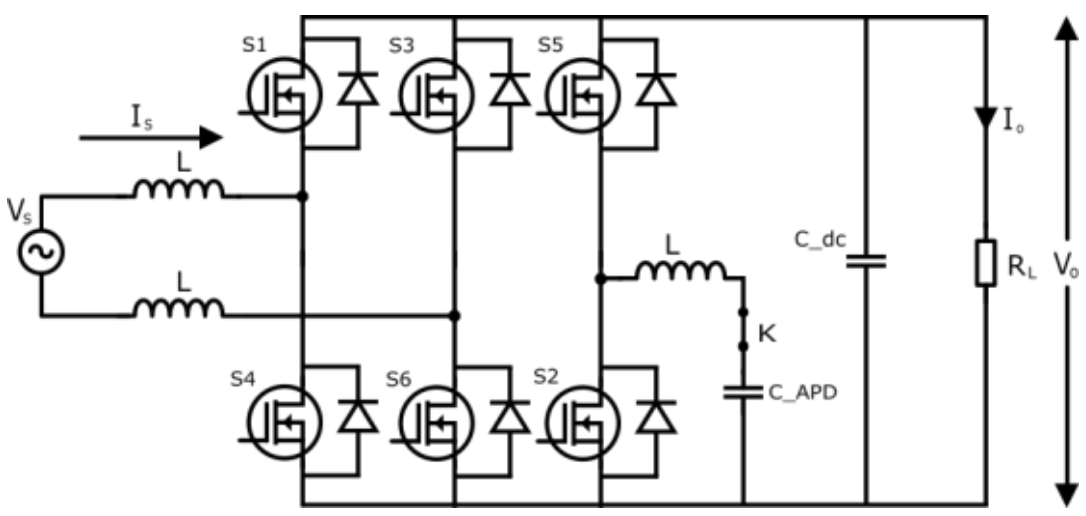

Fig. 3: Equivalent circuit of the proposed converter during single-phase operation.

## III. Operations and Modulation

Both three-phase and single-phase modes along with their modulation strategies are described below.

### A. Three Phase Operation

The proposed three-phase PWM rectifier is a conventional PWM rectifier controlled by sinusoidal PWM (SPWM) with third-harmonic injection, as shown in Fig. 4. The modulation index is selected according to the required AC input voltage and DC output voltage. Based on the desired current ripple and switching frequency, the input-side inductance is designed. Finally, the DC-link capacitor is designed based on the permissible DC-link voltage ripple. The DC-link voltage and current are verified under steady-state conditions.

Formula of modulation index m,

$$m = \frac{V_m}{V_o} \tag{1}$$

Where $V_m$ is the peak input voltage and $V_o$ is the required output voltage.

$D$ is the duty ratio for the boost operation.

$$D = 1 - \frac{V_m}{V_o} \tag{2}$$

For the output capacitance, change in energy because of voltage deviation if the transient peak voltage is $V_{omax}$,

$$\Delta E = \frac{1}{2} C_{dc} (V_o^2 - V_{o_{max}}^2) \tag{3}$$

$$\Delta E = \Delta P t_{max} \tag{4}$$

$$V_o^2 - V_{o_{max}}^2 = (V_o - V_{o_{max}})(V_o + V_{o_{max}}) \tag{5}$$

$$V_o - V_{o_{max}} = \Delta V_o \tag{6}$$

$$V_o + V_{o_{max}} \approx 2V_o \tag{7}$$

Equation is modified to,

$$\Delta E = \Delta P t_{max} = C_{dc} V_o \Delta V_o \tag{8}$$

$$C_{dc} = \frac{\Delta P t_{max}}{V_o \Delta V_o} \tag{9}$$

$$t_{max} = t_{on} = D_{max} T = \frac{D}{f_{sw}} \tag{10}$$

Formula of $C_{dc}$ will be,

$$C_{dc} = \frac{\Delta P D}{f_{sw} V_o \Delta V_o} \tag{11}$$

Most of the values of different components are being selected in the three-phase mode. Based on specified output voltage ripple and predefined grid side inductance, the value Capacitance can be calculated. The capacitor $C_{dc}$ is the same capacitor $C_{3ph}$ in the Fig. 4.

*B. Single phase rectifier without Active Power Decoupling Circuit*

The single-phase operating mode uses the same hardware as the three-phase rectifier. It is also operated by Sinusoidal PWM. The modulation index is selected according to the required AC output voltage. Based on the desired current ripple and switching frequency, the inductance is being selected in three phase mode but the same physical circuit is working as a single phase rectifier, therefore we can't change inductance here. Since the two input inductors are effectively connected in series from the single-phase source perspective, the equivalent input inductance becomes approximately twice the inductance of one phase.

The single-phase operating mode uses the same hardware as the three-phase rectifier. It is also operated by Sinusoidal PWM. Now, The DC-link capacitor that is used in three-phase operation stays connected to the output because the same physical converter works in both operating modes. However the amount of capacitance needed to keep the DC-link voltage ripple, in single-phase operation is much larger because the power pulsation occurs at twice the line frequency. The DClink capacitor used for three phase rectifier is also connected at the output but it has been seen that the DC-link capacitance for the single-phase rectifier is generally very bulky so it won't affect the new output.

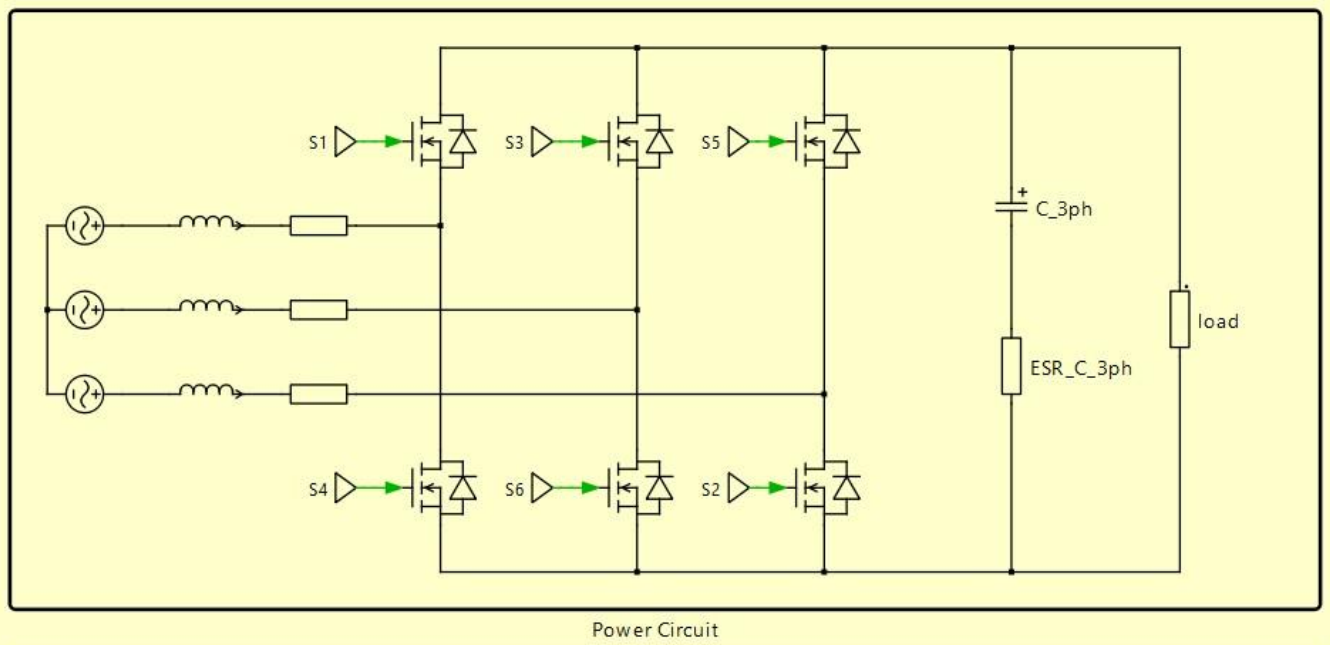

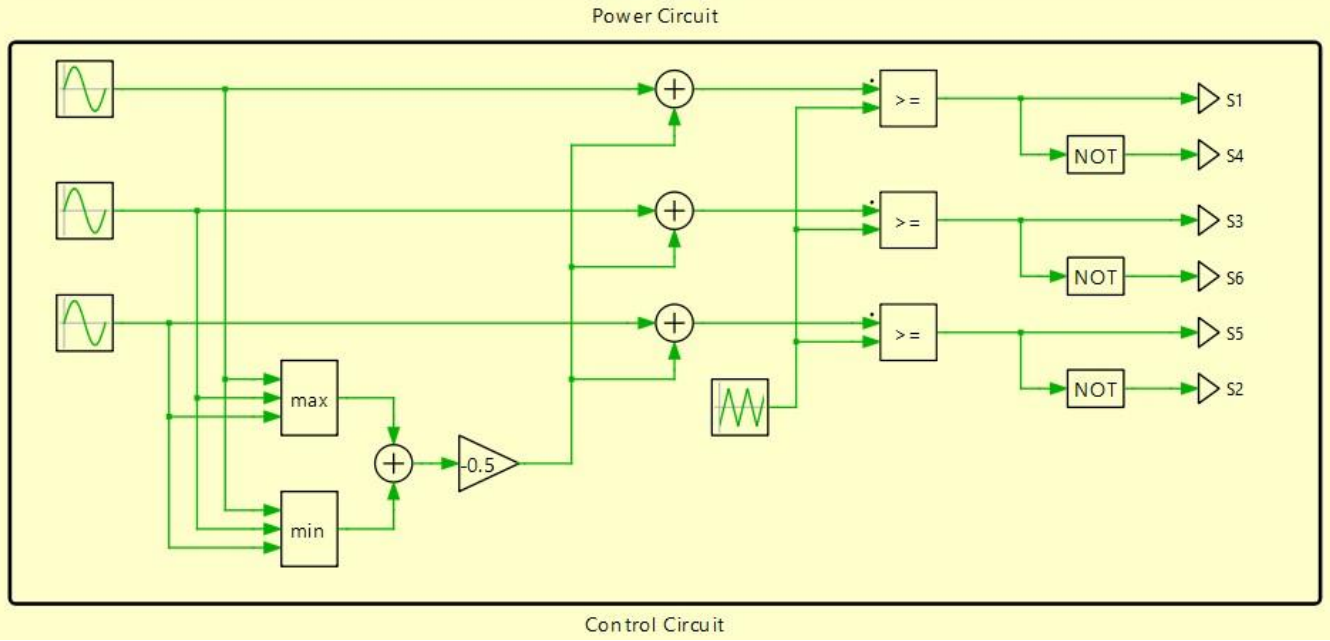

Fig. 4: Three phase rectifier and the Control scheme for three phase operation.

With unity power factor the instantaneous power equation will be,

$$P_r(t) = -(\frac{V_m I_m}{2} cos(2\omega t) + \frac{\omega L I_m^2}{2} sin(2\omega t)) \tag{12}$$

$P_r(t)$ is the instantaneous ripple power [9] at the output of the converter. In the equation 13 the capacitance formula is also given [9]. If the input voltage and input current are assumed to be sinusoidal with unity power factor, the formula of capacitor become,

$$C_{dc} = \frac{\sqrt{P_o^2 + (\frac{2\omega L P_o^2}{V_m^2})^2}}{2\omega V_o \Delta V_o}; \tag{13}$$

Using this formula and the output voltage ripple specification we can calculate the minimum bus-capacitance. Generally this capacitance is very bulky. The required capacitance can be significantly reduced by using an active power decoupling circuit. In single phase rectifier, the problem of double line frequency power at the output should be reduced. As our circuit is same as the three phase circuit and one of the leg is unused till now, that leg will become the Active Power Decoupling (APD) circuit.

*C. Single phase rectifier with Active Power Decoupling Circuit*

To avoid using a large DC-link capacitor for absorbing this second-harmonic energy, the unused third converter leg is utilized as an active power decoupling circuit.The APD circuit operates bidirectionally to buffer the double-line-frequency energy. Whenever the instantaneous input power goes above the power that the load receives the excess energy is transferred to the auxiliary capacitor. In the interval the stored energy is returned to the DC link. Therefore, the main DC-link capacitor primarily supports the average-power transfer and maintains the DC-link voltage, while the auxiliary capacitor absorbs and releases the double-line-frequency ripple energy. The APD is designed to operate in CCM for which the same 1mH inductance is used. A relay is connected between the inductance and auxiliary capacitance to change over between two different operating mode. The proposed auxiliary circuit is shown in Fig.5.

Ripple energy formula is,

$$E_r = \frac{1}{2} C_{APD} V_{cs,max}^2 \tag{14}$$

Minimum capacitance is,

$$C_{APD,min} = \frac{2E_r}{V_{cs,max}^2} \tag{15}$$

We can also write,

$$E_r = \frac{1}{2} C_{APD} (V_{cs,max}^2 - V_{cs,min}^2) \tag{16}$$

$V_{cs,max}$ & $V_{cs,min}$ are the maximum and minimum capacitance of the APD capacitor. From (14)–(16), the minimum auxiliary capacitance can be determined. A practically available capacitor value is selected for this design. The duty ratio of the third leg is then determined from the APD capacitor's voltage equations.

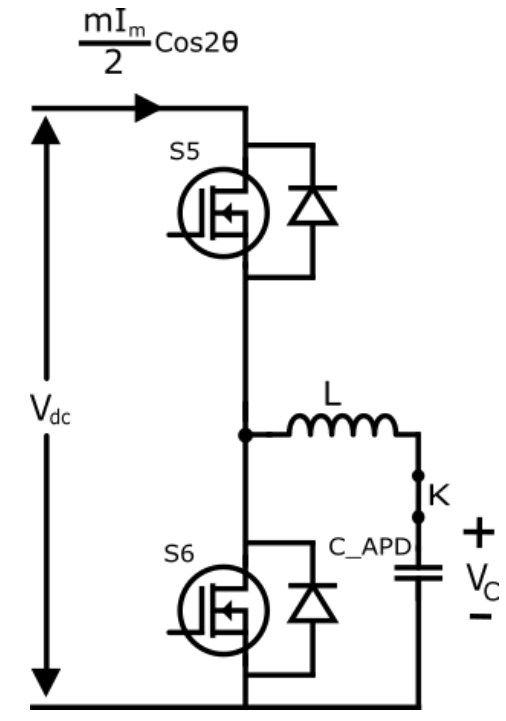

Fig. 5: Single-Phase Equivalent circuit

To determine the duty ratio of the third converter leg, the capacitor-voltage dynamics of the APD stage are considered. The relationship between the capacitor voltage, converter modulation and ripple power is given by

$$C_{APD} \frac{dV_c}{dt} V_c = \frac{mI_m}{2} cos(2\omega t + \phi) V_o \tag{17}$$

Here, $V_{\max}$ denotes the maximum voltage of the APD capacitor, and $V_{dc}$ denotes the nominal DC-link voltage.

$$V_c{}^2 = V_{max}^2 + \frac{mI_m V_o}{2\omega C_{APD}} [sin(2\omega t + \phi) - 1] \tag{18}$$

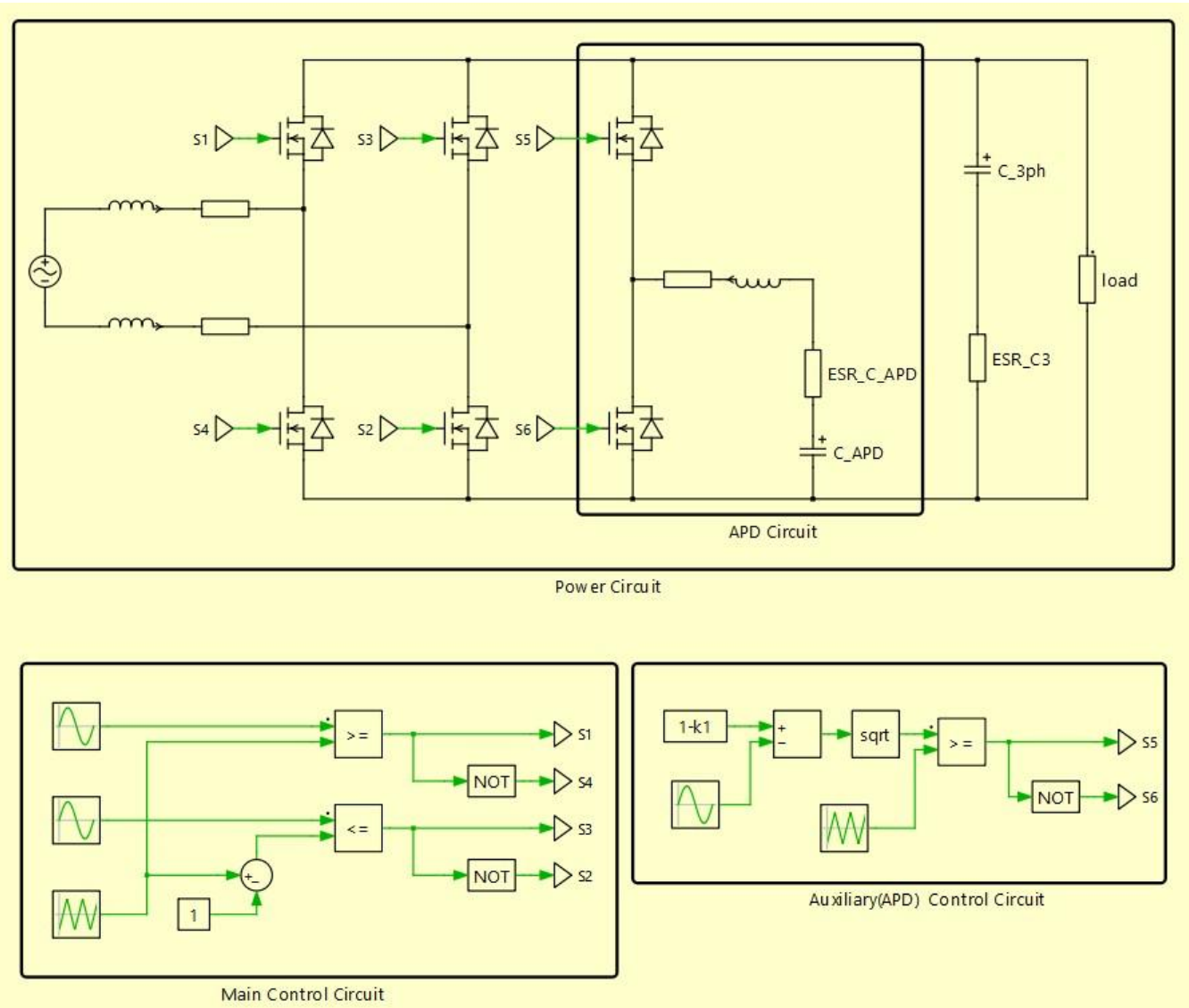

Fig. 6: Single-phase equivalent circuit with the APD branch and corresponding control circuits

Here

$$V_{max} = V_o \tag{19}$$

$$D = \frac{V_c}{V_o} \tag{20}$$

Using equation 18, 19 & 20, the duty cycle for S5 is,

$$D_5 = \sqrt{1 - \frac{mI_m V_o}{2\omega C_{APD}} [sin(2\omega t + \phi) - 1]} \tag{21}$$

and $D_6$ is,

$$D_6 = 1 - D_5 \tag{22}$$

## IV. Simulations

The simulations of the proposed topology is performed using PLECS simulation software. The specifications of the converter are given in table 1.

TABLE I: Design Parameters and Ratings

| Parameters | Specification |
|---|---|
| Three phase grid voltage, $V_{LL}$ | 400V rms |
| Single phase grid Voltage, $V_{in}$ | 230V rms |

| Three phase rated power, $P_{3\phi}$ | 10kW |
|---|---|
| Single phase rated power, $P_{1\phi}$ | 3.3kW |
| Output voltage, $V_o$ | 600V |
| Line inductance, $L_{in}$ | 1mH |
| Switching Frequency | 50kHz |
| DC-Link ripple $\Delta V_{out}$ | 1% |
| Allowable THD | ≤5% |

### *A. Three phase rectifier*

To meet the specified 1% output voltage ripple with the 1mH grid side inductor, the required output bus-capacitance is 26$\mu$F. A 36 Ω load resistance is connected at the output to draw rated power. Fig. 7(a) shows the voltage, current and power at the output side of the converter. Fig. 7(b) shows the input three phase voltages and currents. The voltage stress and the current stress of a switch is shown in Fig. 8. The THD of the input current is 1.96%, while the unity power factor operation is verified by the simulation results.

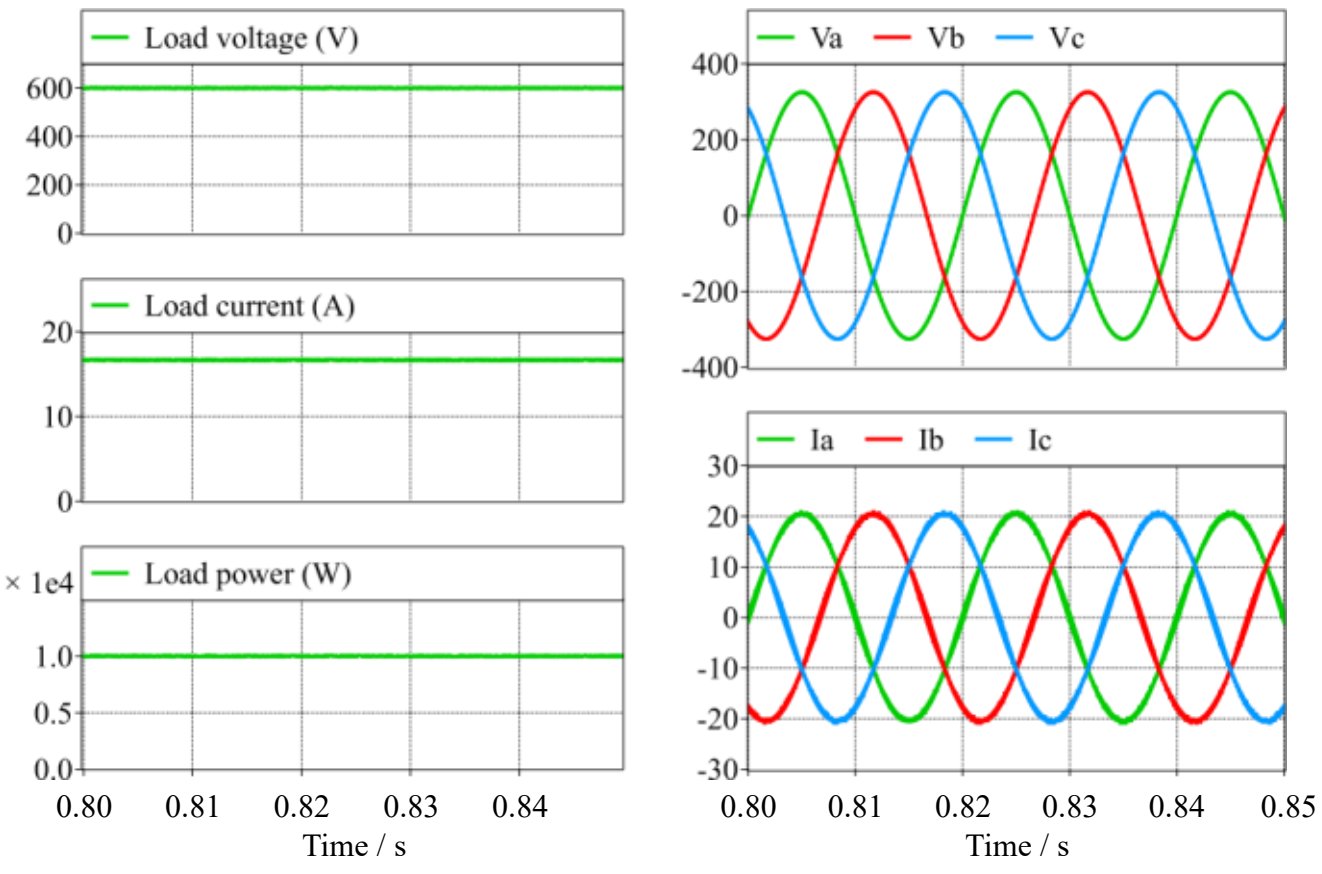

(a) Regulated DC-link Voltage, load (b) Three phase Voltage & Current Current & load Power

Fig. 7: Simulated output power, output and input voltage/current waveforms during three-phase operation

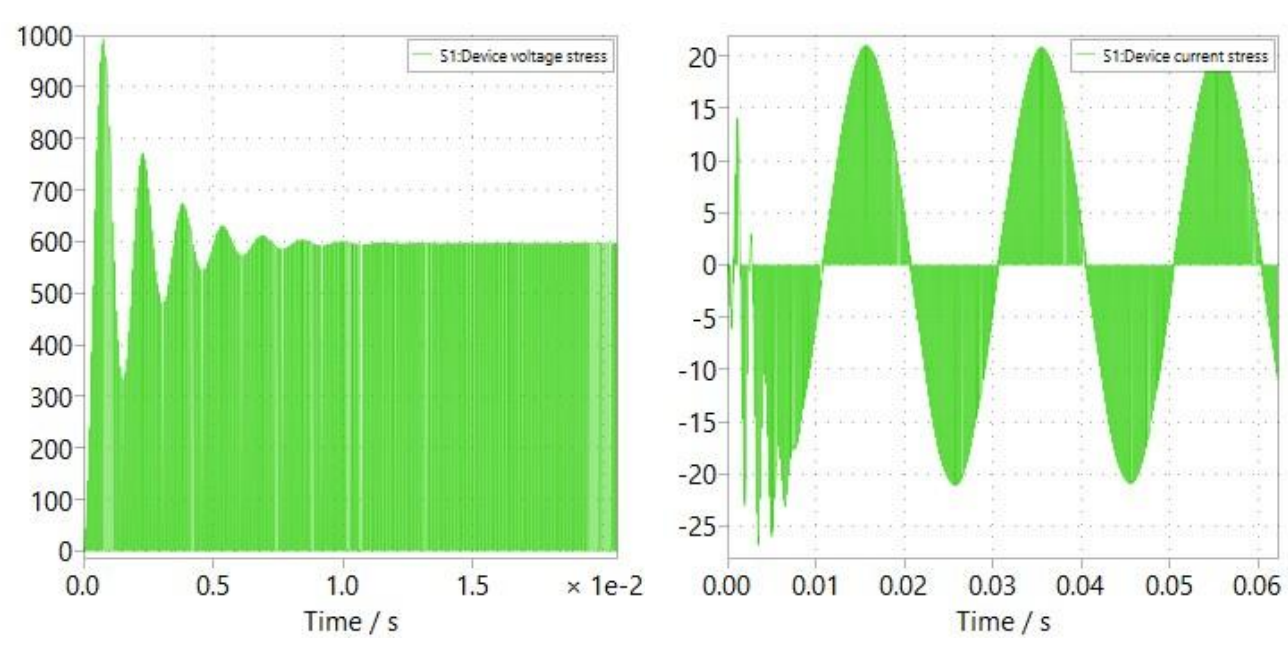

(a) Voltage stress across one Switch (b) Current stress of one switch

Fig. 8: Voltage and Current stresses during three-phase operation

### *B. Single phase rectifier without APD*

When operated in single phase mode with the same specifications, the minimum bus capacitance is 3mF, which is a bulky DC-link capacitor. The third leg of the converter is also unused. Again, to get the specific power a 109 Ω resistance is connected at the output. The resulting transient response is considerably slower, causing the output power to take a longer time to reach its steady-state value. Using APD circuit we can reduce this big bus-capacitance to a significantly low capacitance. Fig. 9 shows the input single phase voltages and currents, which verifies unity power factor operation. The output voltage waveform is shown in Fig. 10. The observed THD in input current is 3.5% and output voltage ripple is 6v.

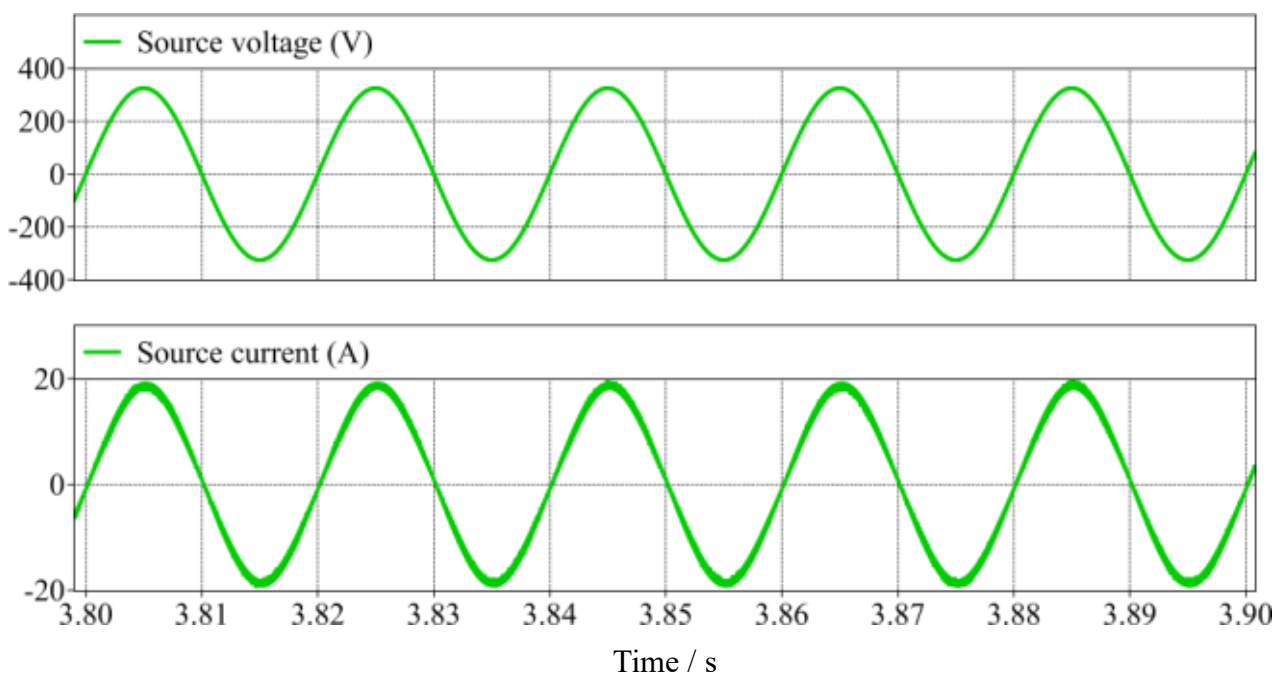

Fig. 9: Simulated single-phase input voltage and current without APD operation ($C_{dc} = 3mF$, THD=3.5%).

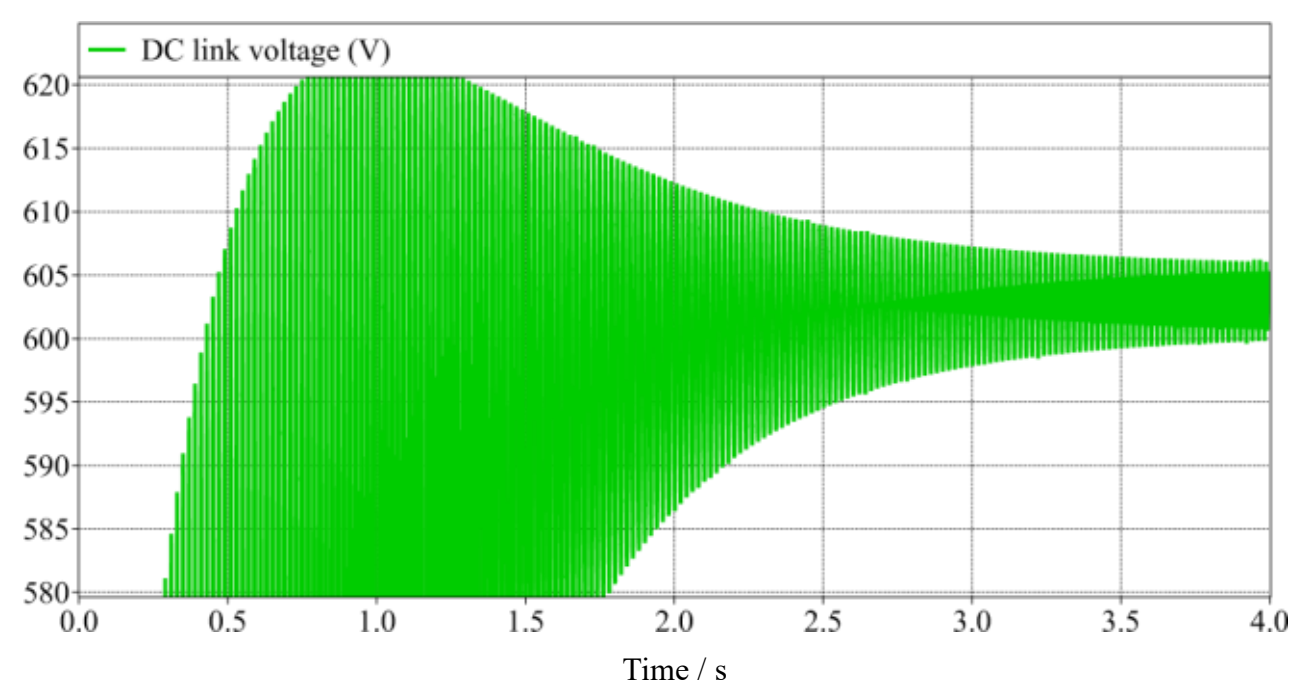

Fig. 10: Simulated output voltage without APD operation ($C_{dc} = 3mF$, $\Delta V_o$=6V).

### *C. Single phase rectifier with APD*

The APD capacitance is 82$\mu$F and the dc link capacitor is 100$\mu$F. Load resistance is 109 Ω. The settling time is very less compared to a single phase rectifier without APD circuit. Fig. 11 shows the input single phase voltages and currents, which verifies unity power factor operation. The output voltage waveform is shown in Fig. 12. The observed THD in input current is 3.6% and output voltage ripple is 6v. Voltage stress and current stress of any device is being shown in Fig. 13.

Without APD the output power response is very slow which is being verified by the simulation. The output power requires approximately 4 s to reach its steady-state value without APD, but with the APD, configuration reaches steady state significantly faster. The upper curve of Fig. 14 is showing the output power of the converter without APD. With APD, the response is very fast, which is being shown in bottom of Fig. 14 .

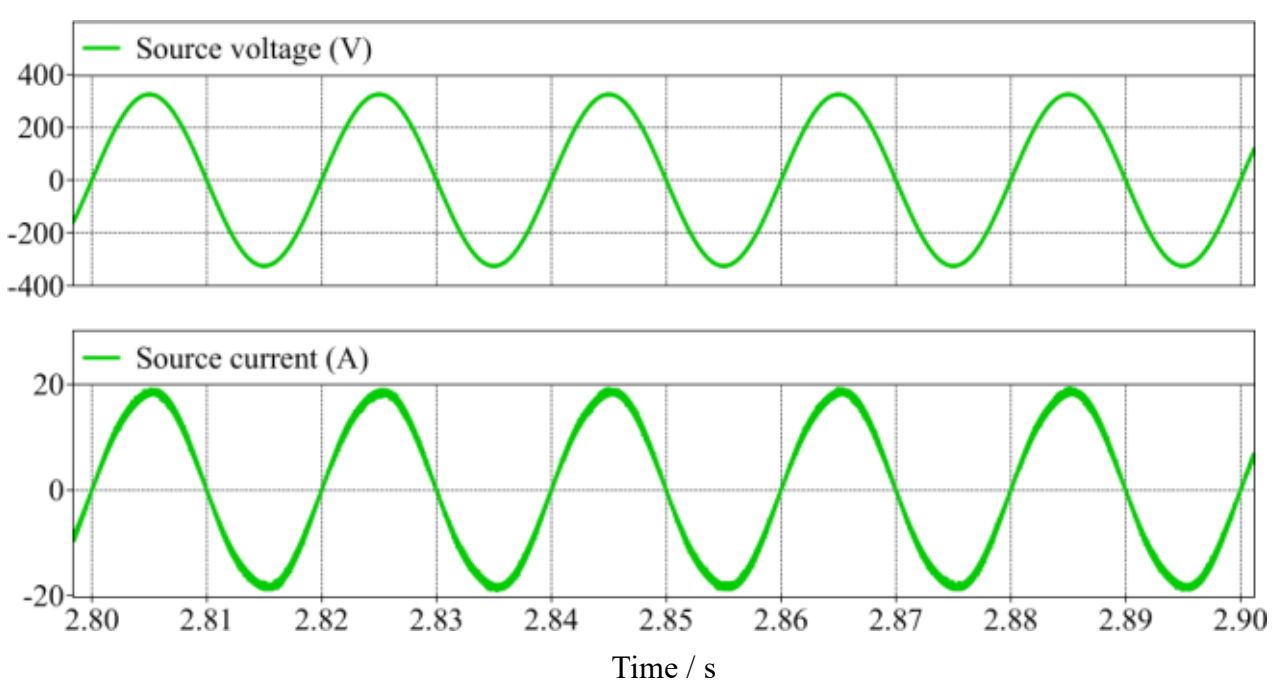

Fig. 11: Simulated single-phase input voltage and current during APD operation ($C_{dc} = 100\mu F$, $C_{APD} = 82\mu F$, THD=3.6%).

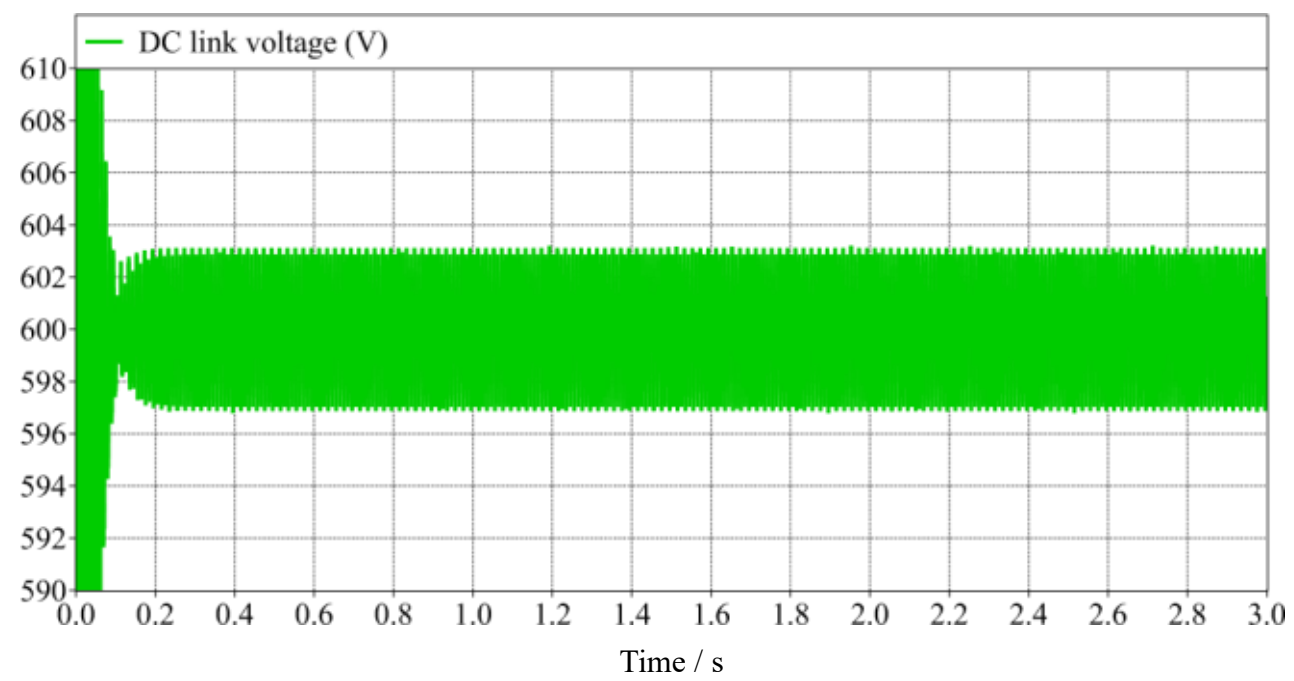

Fig. 12: Simulated output voltage during APD operation ($C_{dc} = 100\mu F$, $C_{APD} = 82\mu F$, $\Delta V_o$=6V).

## V. Conclusions

A new OBC topology is proposed where both three phase and single phase charging is possible with a relay switch and a small capacitor added to a conventional three-phase rectifier. The additional circuit elements along with the redundant third leg achieves active power decoupling to avoid the bulky DC link capacitor during single phase charging. The converter operation, modulation strategy, and parameter design has bee explained. The OBC is designed to supply 10 kW during three phase charging and 3.3 kW during single phase charging to a dc bus of 600V. The simulation model is developed in PLECs to verify converter operation and control strategy. The simulation results show that the performance criteria of 1% output voltage ripple and maximum 5% input current THD can be met using only a 100$\mu$F DC link capacitance and 82$\mu$F APD capacitance. This capacitance is drastically lower than 3 mF DC link capacitance requirement if APD is not available to meet the same performance criteria. Thus, the proposed OBC offers significantly smaller dual mode OBC for EVs.

## References

[1] J. Yuan, L. Dorn-Gomba, A. D. Callegaro, J. Reimers, and A. Emadi, "A review of bidirectional on-board chargers for electric vehicles," *IEEE Access*, vol. 9, pp. 51501–51518, 2021.

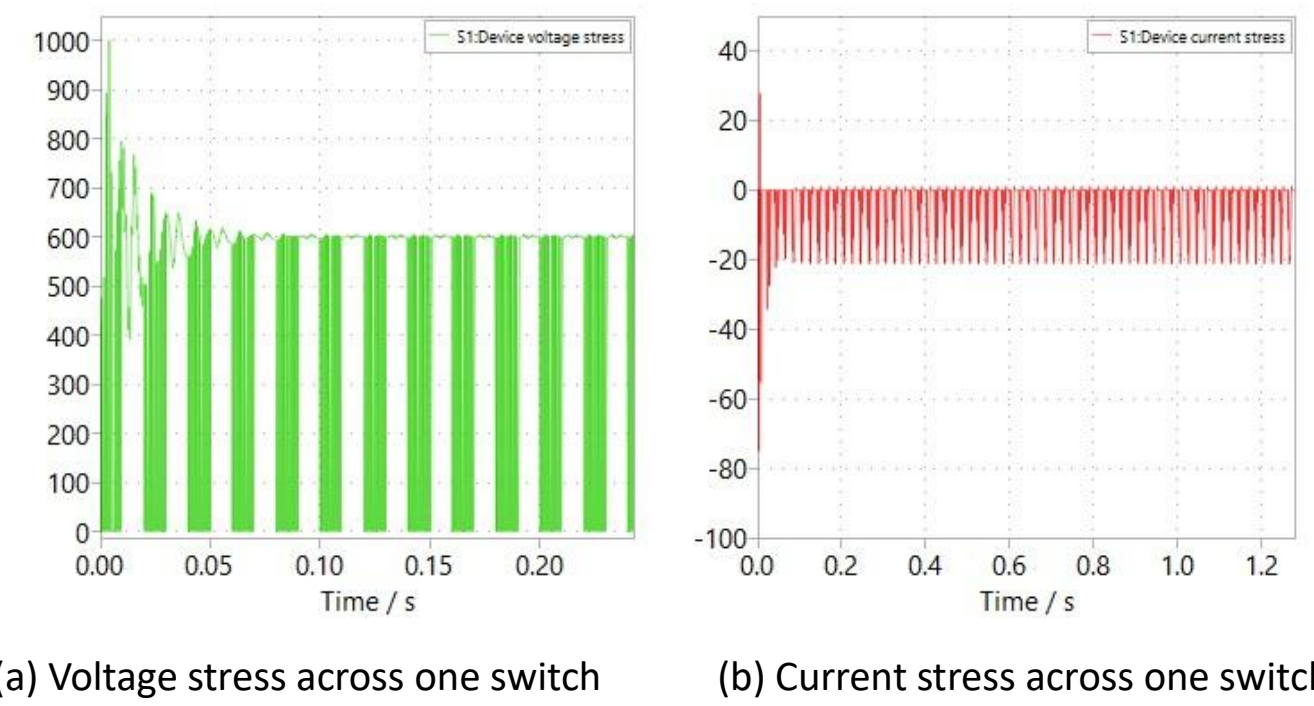

(a) Voltage stress across one switch (b) Current stress across one switch

Fig. 13: Voltage and current stress of a converter switch during single-phase APD operation

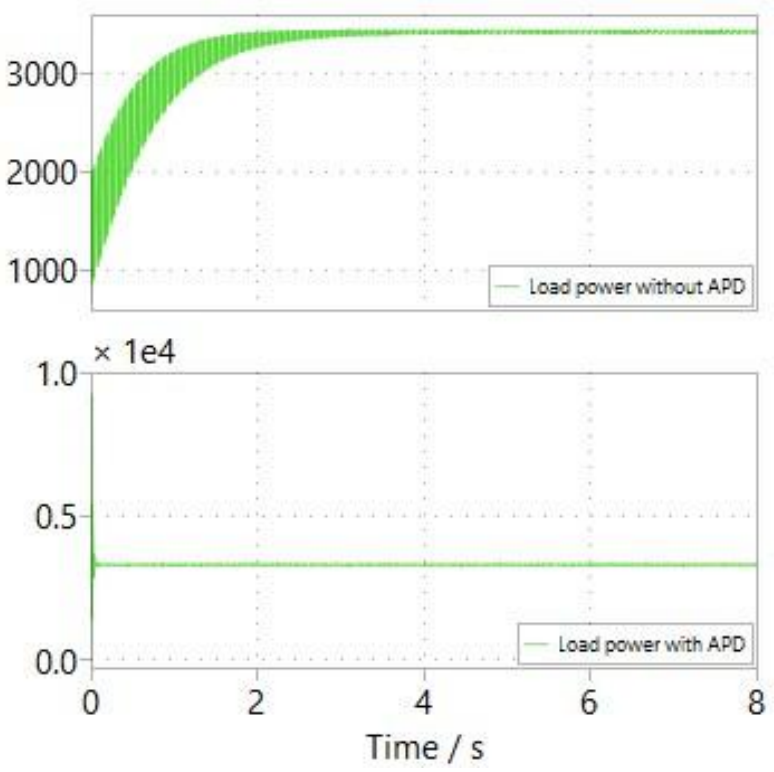

Fig. 14: Comparison of output-power response without & with APD

[2] K. Zhou, H. Yang, Y. Zhang, Y. Che, Y. Huang, and X. Li, "A review of the latest research on the topological structure and control strategies of on-board charging systems for electric vehicles," *Journal of Energy Storage*, vol. 97, p. 112820, 2024.

[3] Z. Wang, X. Su, N. Zeng, and J. Jiang, "Overview of isolated bidirectional dc–dc converter topology and switching strategies for electric vehicle applications," *Energies*, vol. 17, no. 10, 2024.

[4] P. Fang, B. Sheng, W.-B. Liu, Y.-F. Liu, and P. C. Sen, "Parallel energy buffering led driver achieves electrolytic capacitor-less and flicker-free operation," in *2018 IEEE Energy Conversion Congress and Exposition (ECCE)*, pp. 5117–5124, 2018.

[5] H. Sarnago, O. Luc´ ´ıa, D. Menzi, and J. W. Kolar, "Single-/three-phase bidirectional ev on-board charger featuring full power/voltage range and cost-effective implementation," in *2023 IEEE 17th International Conference on Compatibility, Power Electronics and Power Engineering (CPE-POWERENG)*, pp. 1–6, IEEE, 2023.

[6] J. Kolar, U. Drofenik, and F. Zach, "Vienna rectifier ii-a novel singlestage high-frequency isolated three-phase pwm rectifier system," *IEEE Transactions on Industrial Electronics*, vol. 46, no. 4, pp. 674–691, 1999.

[7] Y. Sun, Y. Liu, M. Su, W. Xiong, and J. Yang, “Review of active power decoupling topologies in single-phase systems,” *IEEE Transactions on Power Electronics*, vol. 31, no. 7, pp. 4778–4794, 2016.

[8] H. Li, K. Zhang, H. Zhao, S. Fan, and J. Xiong, “Active power decoupling for high-power single-phase pwm rectifiers,” *IEEE Transactions on Power Electronics*, vol. 28, no. 3, pp. 1308–1319, 2013.

[9] R. Wang, F. Wang, D. Boroyevich, R. Burgos, R. Lai, P. Ning, and K. Rajashekara, “A high power density single-phase pwm rectifier with active ripple energy storage,” *IEEE Transactions on Power Electronics*, vol. 26, no. 5, pp. 1430–1443, 2010.